\documentstyle[12pt]{article}

\title{Mass Generation in the Supersymmetric Nambu--Jona--Lasinio 
Model in an External Magnetic Field\thanks{This talk is based on 
the work done in collaboration with V.Elias, D.G.C.McKeon, and 
V.A.Miransky [1].}} 

\author{ I.A.Shovkovy\\
{\sl Bogolyubov Institute for Theoretical Physics}\\
       {\sl 252143 Kiev, Ukraine} } 

\date{January 6, 1996}

\begin{document}

\maketitle

\begin{abstract}
The mass generation in the ($3+1$)--dimensional supersymmetric 
Nambu--Jona--Lasinio model in a constant magnetic field is 
studied. It is shown that the external magnetic field catalyzes 
chiral symmetry breaking.
\end{abstract}

It was shown in \cite{1,2} and later confirmed in 
\cite{Ng,Hong} that a constant magnetic field is a strong 
catalyst of dynamical chiral symmetry breaking, leading to the 
generation of a fermion dynamical mass even at the weakest 
attraction between fermions (the prehistory of the 
question includes \cite{pre} among others).

The effect is accounted for the effective dimensional reduction 
$D\to D-2$ of the infrared dynamics responsible for the fermion 
pairing in a magnetic field. This reduction is a reflection of 
simple physics:  the motion of charged particles is partly 
restricted in the plane perpendicular to the magnetic field.  The 
latter is also related to the fact that the chiral condensate 
mainly appears due to the lowest Landau level whose dynamics is 
($D-2$)--dimensional.  

In this talk I shall briefly present the results for the 
supersymmetric Nambu--Jona--Lasinio (SNJL) model in a magnetic 
field. 

The motivation for the problem is the following. As was 
heuristically proved in \cite{1,2}, the catalysis by an external 
magnetic field is a rather universal (model--independent) 
phenomenon. In non--supersymmetric models, chiral symmetry breaking 
is usually realized if the coupling constant is large enough. As 
for the influence of the magnetic field, it reduces the critical 
coupling to zero.  On the other hand, there is no spontaneous 
chiral symmetry breaking in the SNJL model at all \cite{6}. 

Below it will be shown that an external magnetic field changes 
the situation in the SNJL model dramatically: chiral symmetry 
breaking, in agreement with the universality of the effect 
\cite{2}, occurs for any value of the coupling constant. 

The action of the SNJL model with the $U_L(1)\times U_R(1)$ chiral 
symmetry in a magnetic field (in notations of Ref.\cite{13} except 
the metric $g^{\mu\nu}=\mbox{diag}(1,-1,-1,-1)$) is:
\begin{eqnarray}
\Gamma&=&\int d^8z 
    \left[ \bar{Q}e^{V}Q+\bar{Q}^ce^{-V}Q^c 
       + G(\bar{Q}^c\bar{Q})(QQ^c)\right] .
\label{eq33}
\end{eqnarray}
Here $d^8z=d^4xd^2\theta d^2\bar{\theta}$, $Q^{\alpha}$ and 
$Q^c_{\alpha}$ are chiral superfields carrying the color index 
$\alpha=1, 2, \dots, N_c$, i.e.  $Q^{\alpha}$ and $Q^c_{\alpha}$ 
are assigned to the fundamental and antifundamental representations 
of the $SU(N_c)$, respectively:
\begin{eqnarray}
          Q^{\alpha}  = \varphi^{\alpha} 
                      + \sqrt{2}\theta\psi^{\alpha}
                      + \theta^2F^{\alpha} , &\quad&
         Q^c_{\alpha} = \varphi^c_{\alpha}
                      + \sqrt{2}\theta\psi^c_{\alpha}
                      + \theta^2F^c_{\alpha} 
\label{eq34}
\end{eqnarray}
(henceforth I shall omit color indices). The vector superfield 
$V(x,\theta,\bar{\theta}) =-\theta \sigma^{\mu} \bar{\theta} 
A^{ext}_\mu$, with $A^{ext}_\mu = B x^2 \delta_{\mu}^3$, describes 
an external magnetic field in the $+x_1$ direction.

The action (\ref{eq33}) is equivalent to the following one:
\begin{eqnarray}
\Gamma_{A} &=& \int d^8z 
           \left[\bar{Q}e^{V}Q+\bar{Q}^ce^{-V}Q^c
                 + \frac{1}{G}\bar{H}H\right]-
           \nonumber\\
               &&- \int d^6z\left[\frac{1}{G}HS-QQ^cS\right]
                 - \int d^6\bar{z}\left[\frac{1}{G}\bar{H}\bar{S}
                 -                \bar{Q}\bar{Q}^c\bar{S}\right] .
\label{eq35}
\end{eqnarray}
Here 
$d^6z=d^4xd^2\theta$, 
$d^6\bar{z}=d^4xd^2\bar{\theta}$,
and $H$ and $S$ are two auxiliary chiral fields:
\begin{eqnarray}
      H=h+\sqrt{2}\theta\chi_h+\theta^2f_h , &\quad&
      S=s+\sqrt{2}\theta\chi_s+\theta^2f_s .
\label{eq36}
\end{eqnarray}
The Euler--Lagrange equations for these auxiliary fields take the
form of constraints:
\begin{eqnarray}
               H = GQQ^c, &\quad&
               S = - \frac{1}{4}\bar{D}^2(\bar{H})=
                   - \frac{G}{4}\bar{D}^2(\bar{Q}\bar{Q}^c).
\label{eq37}
\end{eqnarray}
Here $\bar{D}$ is a SUSY covariant derivative \cite{13}. The action 
(\ref{eq35}) reproduces Eq.(\ref{eq33}) upon application of the 
constraints (\ref{eq37}).

In terms of the component fields, the action (\ref{eq35}) is
\begin{eqnarray}
\Gamma_{A} &=& \int d^4x \Bigg[
- \varphi^{\dagger}(\partial_{\mu}-ieA^{ext}_{\mu})^2\varphi
- \varphi^{c\dagger}(\partial_{\mu}+ieA^{ext}_{\mu})^2\varphi^c        
\nonumber \\
&&+ i\bar{\psi}\bar{\sigma}^{\mu}
(\partial_{\mu}-ieA^{ext}_{\mu})\psi
+ i\bar{\psi}^c\bar{\sigma}^{\mu}
(\partial_{\mu}+ieA^{ext}_{\mu})\psi^c
+ F^{\dagger}F + F^{c\dagger}F^c 
\nonumber \\
&&+ \frac{1}{G}\left( -h^{\dagger}\Box h 
+ i\bar{\chi}_h\bar{\sigma}^{\mu}\partial_{\mu}\chi_h 
+ f^{\dagger}_hf_h \right)
+ \frac{1}{G}\left( \chi_h\chi_s - hf_s- sf_h + h.c.\right)
\nonumber \\
&&- \Big( s\psi\psi^c + (\varphi\psi^c+\varphi^c\psi)\chi_s
         - s(\varphi F^c + \varphi^cF) - \varphi\varphi^c f_s 
         + h.c.\Big)
         \Bigg] .
\label{eq38}
\end{eqnarray}
To obtain the effective potential, all the auxiliary scalar fields 
are treated as (independent of $x$) constants and all the auxiliary 
fermion fields equal zero.  Then, the Euler--Lagrange equations for 
the fields $F$, $F^c$, $f_h$, $h$ and their conjugates leads to 
$F^{\dagger}=-s\varphi^c$, $F^{c\dagger}=-s\varphi$, 
$f^{\dagger}_h=s$, $f^{\dagger}_s=0$, plus h.c. equations. After 
taking these into account, the action reads
\begin{eqnarray}
\Gamma_{A} &=& \int d^4x \Bigg[
- \varphi^{\dagger}\left[(\partial_{\mu}-ieA^{ext}_{\mu})^2
                    + \rho^2 \right]\varphi
- \varphi^{c\dagger}\left[(\partial_{\mu}+ieA^{ext}_{\mu})^2
                    + \rho^2 \right] \varphi^c        
\nonumber \\
&&+ i\bar{\psi}_D\gamma^{\mu}(\partial_{\mu}
        -ieA^{ext}_{\mu})\psi_D - \sigma\bar{\psi}_D\psi_D
        - \pi\bar{\psi}_Di\gamma^5\psi_D - \frac{\rho^2}{G}
\Bigg] ,
\label{eq39}
\end{eqnarray}
where $s=\sigma+i\pi$, $\rho^2=|s|^2=\sigma^2+\pi^2$, and the Dirac 
fermion field $\psi_D$ is introduced.

In leading order in $1/N_c$, the effective potential $V(\rho)$ can 
now be derived in the same way as in the ordinary NJL model. 
The difference is that, besides fermions, the two scalar 
fields $\varphi^c$ and $\varphi$ give a contribution to $V(\rho)$:
\begin{equation}
V(\rho) = \frac{\rho^2}{G} 
               + V_{fer}(\rho) + 2 V_{bos}(\rho),
\label{eq40}
\end{equation}
where 
\begin{eqnarray}
V_{fer}(\rho) &=& \frac{N_c}{8\pi^2 l^4} 
               \int\limits^\infty_{1/(l\Lambda)^2} 
               \frac{ds}{s^2}\exp\left(-s( l\rho)^2\right) \coth{s},
\label{eq41} \\
V_{bos}(\rho) &=& -\frac{N_c}{16\pi^2 l^4} 
               \int\limits^\infty_{1/(l\Lambda)^2} 
               \frac{ds}{s^2}\exp\left(-s( l\rho)^2\right) 
               \frac{1}{\sinh{s}} .
\label{eq42}
\end{eqnarray}
This can be rewritten as \cite{0}:
\begin{eqnarray}
V(\rho) &=& \frac{N_c}{8\pi^2 l^4} 
        \Bigg[ \frac{( l\rho)^2}{g}
           + ( l\rho)^2\left(1-\ln\frac{( l\rho)^2}{2}\right)
           + 4\cdot\int\limits_{( l\rho)^2/2}^{[( l\rho)^2+1]/2}
             dx \ln\Gamma(x)\Bigg]+
        \nonumber\\
          &+& \frac{N_c}{16\pi^2 l^4} 
        \Bigg[ \ln(\Lambda l)^2-\gamma -\ln(8\pi^2) \Bigg]
            + O\left(\frac{1}{\Lambda}\right),
\label{eq43}
\end{eqnarray}
where the dimensionless coupling constant is $g=GN_c/8\pi^2l^2$.

As the magnetic field $B$ goes to zero ($l\to\infty$), one obtains: 
\begin{equation} 
V(\rho) = \frac{\rho^2}{G} .  
\label{eq44}
\end{equation}
This potential is positive--definite, as has to be in a 
supersymmetric theory. The only minimum of this potential is 
$\rho=0$ what corresponds to the chiral symmetric vacuum \cite{6}.

The presence of a magnetic field changes this situation
dramatically: at $B\neq 0$, a non--trivial global minimum,
corresponding to chiral symmetry breaking, exists for all
$g>0$. 

The gap equation $dV/d\rho=0$, following from Eq.(\ref{eq43}), is
\begin{equation}
\frac{N\rho}{4\pi^2 l^2}
      \Bigg[ \frac{1}{g}-\ln\frac{( l\rho)^2}{2}
          + 2\ln\Gamma\left(\frac{( l\rho)^2+1}{2}\right)
          - 2\ln\Gamma\left(\frac{( l\rho)^2}{2}\right)
      \Bigg] = 0.
\label{eq45}
\end{equation}
It is easy to check that, at $B\neq 0$, the trivial
solution $\rho=0$ corresponds to a maximum of $V$, since 
$d^2V/d\rho^2|_{\rho\to 0} \to -\infty$. Numerical analysis of 
equation (\ref{eq45}) for $g>0$ and $B\neq 0$ shows that there is a 
nontrivial solution $\rho=m_{dyn}$ which is the global 
minimum of the potential. The analytic expression for $m_{dyn}$ can 
be obtained for small $g$ (when $m_{dyn}l\ll 1$) and for very large 
$g$ (when $m_{dyn}l\gg 1$).  In those two cases, the results are: 
\begin{eqnarray}
     \frac{1}{g}&\simeq& -\ln \frac{\pi(\rho l)^2}{2} ,
     \quad g\ll 1;   \label{eq46}  \\
     \frac{1}{g}&\simeq& \frac{1}{2(\rho l)^2} ,       
     \quad g\gg 1,   \label{eq48}
\end{eqnarray}
i.e.
\begin{eqnarray}
m_{dyn}&\simeq& \sqrt{\frac{2|eB|}{\pi}}
  \exp\left[-\frac{4\pi^2}{|eB|N_cG}\right],
\quad g\ll 1; \label{eq47}\\
m_{dyn}&\simeq& \frac{|eB|}{4\pi}\sqrt{GN_c}, 
\quad g\gg 1. \label{eq49}
\end{eqnarray}
At this point, it seems appropriate to note that the infrared 
dynamics in the SNJL model in a magnetic field is actually 
equivalent to that in the ordinary NJL model in a magnetic field as 
soon as the coupling is weak. This follows from direct comparison 
of the effective potentials and the kinetic terms of the models 
\cite{0}.  The physical picture underlying this equivalence is also 
clear.  The spectra of charged free fermions and bosons in a 
magnetic field are essentially different:
\begin{equation} 
E_n(k_1)=\pm\sqrt{m^2+2|eB|n+k_1^2} , 
\qquad n=0,1,2,\dots  , \label{eq52} 
\end{equation} 
and
\begin{equation} 
E_n(k_1)=\pm\sqrt{m^2+|eB|(2n+1)+k_1^2} , 
\qquad n=0,1,2,\dots .  \label{eq53} 
\end{equation} 
for fermions and bosons, respectively. The crucial difference 
between them is the existence of the gap $\Delta E = \sqrt{|eB|}$ 
in the spectrum of massless bosons and the absence of any gap in 
the spectrum of massless fermions.  Thus at weak coupling, the 
infrared dynamics, responsible for chiral symmetry breaking, is 
dominated by fermions while bosonic degrees of freedom are 
irrelevant. So, it is not a surprise that the infrared dynamics in 
the SNJL and NJL models are equivalent.

In conclusion, I note that the results obtained here are in 
agreement with the general conclusion of Refs.\cite{1,2}, saying 
that the catalysis of chiral symmetry breaking by a magnetic field 
is a universal, model independent effect. 

\section*{Acknowledgments}
I would like to thank the organizers of the Seminar for financial 
support and for the opportunity to give this talk.
I thank V.~Elias, D.G.C.~McKeon, and V.A.~Miransky for enjoyable
collaboration.

\end{document}